\documentclass[12pt]{article}
\usepackage{a4wide}
\usepackage{latexsym}
\usepackage{amsmath}
\usepackage{amsfonts}
\usepackage{amscd}
\usepackage{cite}

\usepackage{pslatex}
\usepackage{graphicx}
\usepackage[latin1]{inputenc}
\usepackage[T1]{fontenc}

\newcommand{\bq}{\begin{eqnarray}}
\newcommand{\eq}{\end{eqnarray}}
\newcommand{\eps}{\varepsilon}

\begin{document}

\thispagestyle{empty}

\begin{flushright}
  MZ-TH/12-21
\end{flushright}

\vspace{1.5cm}

\begin{center}
  {\Large\bf Random polarisations of the dipoles\\
  }
  \vspace{1cm}
  {\large Daniel G\"otz, Christopher Schwan and Stefan Weinzierl\\
\vspace{2mm}
      {\small \em Institut f{\"u}r Physik, Universit{\"a}t Mainz,}\\
      {\small \em D - 55099 Mainz, Germany}\\
  } 
\end{center}

\vspace{2cm}

\begin{abstract}\noindent
  {
We extend the dipole formalism for massless and massive partons
to random polarisations of the external partons.
The dipole formalism was originally formulated for spin-summed matrix elements
and later extended to individual helicity eigenstates.
For efficiency reasons one wants to replace the spin sum by a smooth integration over
additional variables.
This requires the extension of the dipole formalism to random polarisations.
In this paper we derive the modified subtraction terms.
We only modify the real subtraction terms, the integrated subtraction terms do not require any modifications.
   }
\end{abstract}

\vspace*{\fill}

\newpage

\section{Introduction}
\label{sec:intro}

The recent years have witnessed significant progress in the field of next-to-leading order (NLO) calculations.
In particular through new techniques for the computation of the virtual one-loop corrections
a description of multi-parton final states at NLO is no longer out of reach.
Multi-jet final states are a typical signature in the current LHC experiments and a calculation
at NLO is mandatory for an accurate and precise description of these signatures.

The improvements of the techniques for the computation of the virtual one-loop corrections occured along
three lines:
First of all, the traditional approach based on Feynman graphs has been brought 
to perfection \cite{Dittmaier:2003bc,Denner:2002ii,Denner:2005nn,Denner:2010tr,Giele:2004iy,vanHameren:2005ed,Ellis:2005zh,delAguila:2004nf,Pittau:2004bc,Binoth:2002xh,Binoth:2005ff,vanHameren:2009vq,Cascioli:2011va}.
Secondly, techniques based on unitarity or cuts have made a major contribution to the field \cite{Bern:1995cg,Berger:2008sj,Forde:2007mi,Ossola:2006us,Ossola:2007ax,Ossola:2008xq,Mastrolia:2008jb,Draggiotis:2009yb,Garzelli:2009is,Kilgore:2007qr,Anastasiou:2006jv,Anastasiou:2006gt,Ellis:2007br,Giele:2008ve,Ellis:2008ir,Badger:2010nx,Cullen:2011ac,Hirschi:2011pa}.
Finally purely numerical methods for the computation of the virtual one-loop part were shown to be competitive in the multi-parton domain \cite{Becker:2012aq,Becker:2010ng,Assadsolimani:2010ka,Assadsolimani:2009cz,Nagy:2003qn,Gong:2008ww,Nagy:2006xy}.
With the help of these techniques several multi-parton processes have been computed \cite{Berger:2009zg,Berger:2009ep,Berger:2010vm,Berger:2010zx,Ita:2011wn,Bern:2011ep,Ellis:2009zw,KeithEllis:2009bu,Melia:2010bm,Bevilacqua:2010ve,Bevilacqua:2009zn,Frederix:2010ne,Bredenstein:2009aj,Becker:2011vg}.

In view of these improvements for the computation of the virtual one-loop corrections one has to reconsider potential
bottle-necks which hinder us from going to even higher parton multiplicities.
In this re-evaluation the real contribution comes back into focus.
For a NLO calculation the real contribution is given by the Born amplitude squared with one additional parton.
The integration over the phase space of one unresolved parton is divergent.
To render this integration finite, the subtraction method is usually employed \cite{Kunszt:1994mc,Frixione:1996ms,Catani:1997vz,Catani:1997vzerr,Dittmaier:1999mb,Phaf:2001gc,Catani:2002hc,Frederix:2009yq,Frixione:2011kh,Kosower:1998zr,Campbell:1998nn,Gehrmann-DeRidder:2005cm,Daleo:2006xa,Somogyi:2006cz,Nagy:2007ty,Chung:2010fx}.
In current calculations it turns out that the subtracted real contribution is the most challenging one, not from a conceptional point of view, but in terms
of CPU time.

Modern calculations are often based on helicity amplitudes. The use of helicity amplitudes avoids the $N_{\mathrm{diagram}}^2$-scaling of the squared Born amplitude, which
would occur if the Born amplitude is expressed in terms of $N_{\mathrm{diagram}}$ Feynman diagrams.
Instead one sums over all helicity configurations explicitly. 
By using explicit representations for the polarisation vectors and the polarisation spinors the amplitude can be calculated as a complex number for any given
helicity configuration.
Taking the norm of a complex number is a simple operation and negligible in terms of CPU time.
This reduces the complexity from $N_{\mathrm{diagram}}^2$ down to $N_{\mathrm{diagram}}$.
Furthermore, the $N_{\mathrm{diagram}}$-scaling can be avoided by calculating the amplitude through recurrence relations \cite{Berends:1987me}.
Let us now discuss the costs of the helicity method:
Both quarks and gluons have two spin states.
It follows that for an amplitude with $n$ external partons we have a sum over $2^n$ helicity configurations.
It is well known that for the Born contribution and the virtual contribution the sum over $2^n$ helicity configurations
can be replaced by an integration over $n$ random polarisations.
This integration can be done by Monte Carlo methods and can be combined with the Monte Carlo integration over the phase space of the
final state particles.
Doing so will speed up the evaluation of the integrand by a factor of up to $2^n$.
(The speed up can be less than $2^n$, if intermediate results from one helicity configuration are cached and re-used for another helicity configuration.)
As a slight disadvantage we have to take into account that we integrate now over more dimensions (phase space plus $n$ helicity variables) and that
the integrand fluctuates more strongly.
This implies that a larger number of integrand evaluations is needed to reach a prescribed Monte Carlo integration error.
Despite this slight draw-back it turns out empirically that there is a net gain by using the Monte Carlo integration over the random polarisations as compared to summing over all
helicity configurations.

It is therefore natural to extend this method to the real contribution.
Here, a technical problem has to be circumvented:
As mentioned above we need subtraction terms in order to render the integration over the phase space of one unresolved particle
finite.
The subtraction terms are constructed to match locally the singular behaviour of the integrand.
Currently known subtraction terms, like the ones within the dipole formalism \cite{Catani:1997vz,Catani:1997vzerr,Dittmaier:1999mb,Phaf:2001gc,Catani:2002hc},
match locally the behaviour of the spin-summed amplitude squared.
However, they do not match locally the singular behaviour of a squared amplitude with random polarisations.
In this case, additional subtractions terms are needed.
In this paper we provide these additional subtraction terms.
The additional subtraction terms, which we derive, have the following properties:
\begin{enumerate}
\item The additional subtraction terms integrate to zero. This implies that we only need to modify the unintegrated subtraction terms, which
are subtracted from the Born amplitude squared with $(n+1)$ partons.
The integrated form of the subtraction terms, which is added back, is not modified.
\item The additional subtraction terms are independent of the subtraction scheme which is used for the spin-summed squared amplitude.
They can be used with any subtraction scheme, which provides local subtraction terms for the spin-summed squared amplitude.
In this context we remark that the currently known antenna subtraction terms do not provide local subtraction terms for the spin-summed squared amplitude.
The antenna subtraction terms approximate only the angular average of spin-summed squared amplitudes and are thus not local. 
Antenna subtraction has to be used with an additional slicing parameter.
\item The additional subtraction terms are formulated in such a way, that they are independent of the actual definition of the polarisation vectors and polarisation
spinors. We give a representation for the polarisation vectors and polarisation spinors in the appendix, but the reader may use his own definition.
\end{enumerate}
We would like to point out that recently local subtraction terms for squared amplitudes with helicity eigenstates have been 
derived \cite{Czakon:2009ss,Dittmaier:2008md}.
These subtraction terms are however not sufficient for the method of random polarisations. This point will be discussed in detail
in section~\ref{sec:framework}.

This paper is organised as follows: In the next section we introduce the framework.
We review the subtraction method and the method of random polarisations.
In section~\ref{sec:limits} we discuss the soft and collinear limits in the case of random polarisations.
Section~\ref{sec:subtraction_terms} contains the main results of this paper and gives the additional subtraction terms needed to use
the method of random polarisations for the real contribution.
Finally, section~\ref{sec:conclusions} contains our conclusions.
In an appendix we give a choice for the polarisation vectors and polarisation spinors.

\section{The framework}
\label{sec:framework}

The starting point for the calculation of an infrared safe observable $O$ in
hadron-hadron collisions is the following formula:
\bq
\langle O \rangle & = & \sum\limits_{a,b} \int dx_1 f_a(x_1) \int dx_2 f_b(x_2) 
             \frac{1}{2 K(\hat{s}) n_{\mathrm{spin}}(1) n_{\mathrm{spin}}(2) n_{\mathrm{colour}}(1) n_{\mathrm{colour}}(2)}
 \\
 & &
             \int d\phi_{n}\left(p_1,p_2;p_3,...,p_{n+2}\right)
             O\left(p_1,...,p_{n+2}\right)
             \left| {\cal A}_{n+2} \right|^2.
\eq
This equation gives the contribution from the $n$-parton final state.
The two incoming particles are labelled $p_1$ and $p_2$, while $p_3$ to $p_{n+2}$ denote
the final state particles.
$f_a(x)$ gives the probability of finding a parton $a$ with momentum fraction $x$ inside
the parent hadron $h$.
$2K(\hat{s})$ is the flux factor, for massless partons it is given by $2K(\hat{s})=2\hat{s}$.
The quantity $n_{\mathrm{spin}}(i)$ denotes the number of spin degrees of freedom of the parton $i$ and equals
two for quarks and gluons.
Correspondingly, $n_{\mathrm{colour}}(i)$ denotes the number of colour degrees of freedom of the parton $i$.
For quarks, this number equals three, while for gluons we have eight colour degrees of freedom.
Dividing by the appropriate number of degrees of freedom in the initial state
corresponds to an averaging.
$d\phi_n$ is the phase space measure for $n$ final state particles, including (if appropriate) the identical particle factors.
The matrix element $| {\cal A}_{n+2} |^2$ is calculated perturbatively.

The contributions at leading and next-to-leading order are written as
\bq
\langle O \rangle^{LO} & = & 
 \int\limits_n O_n d\sigma^B,
 \nonumber \\ 
\langle O \rangle^{NLO} & = & 
 \int\limits_{n+1} O_{n+1} d\sigma^R + \int\limits_n O_n d\sigma^V 
 + \int\limits_n O_n d\sigma^C.
\eq
Here a rather condensed notation is used. $d\sigma^B$ denotes the Born
contribution,
whose matrix elements are given by the square of the Born amplitudes with $(n+2)$ partons
$| {\cal A}^{(0)}_{n+2} |^2$.
Similarly, $d\sigma^R$ denotes the real emission contribution,
whose matrix elements are given by the square of the Born amplitudes with $(n+3)$ partons
$| {\cal A}^{(0)}_{n+3} |^2$.
$d\sigma^V$ gives the virtual contribution, whose matrix element is given by the interference term
of the one-loop amplitude ${\cal A}^{(1)}_{n+2}$ with $(n+2)$ partons with the corresponding
Born amplitude ${\cal A}^{(0)}_{n+2}$.
$d\sigma^C$ denotes a collinear subtraction term, which subtracts the initial-state collinear
singularities.
Taken separately, the individual contributions are divergent and only their sum is finite.
In order to render the individual contributions finite, such that the phase space integrations
can be performed by Monte Carlo methods, one adds and subtracts a suitably chosen piece
\cite{Catani:1997vz,Catani:1997vzerr,Dittmaier:1999mb,Phaf:2001gc,Catani:2002hc}:
\bq
\langle O \rangle^{NLO} & = & 
 \int\limits_{n+1} \left( O_{n+1} d\sigma^R - O_n d\sigma^A \right)
 + \int\limits_n \left( O_n d\sigma^V + O_n d\sigma^C + O_n \int\limits_1 d\sigma^A \right).
\eq
The squared matrix elements involve a sum over the spins of the external particles.
Since both quarks and gluons have two independent spin states, we can label
the independent spin states of QCD partons by ``$+$'' and ``$-$''.
We thus have for a tree-level amplitude with $n$ external partons
\bq
 \left| {\cal A}^{(0)}_n\left(p_1,...,p_n \right) \right|^2
 & = &
 \sum\limits_{\lambda_1,...,\lambda_n} \left| {\cal A}^{(0)}_n\left(p_1,\lambda_1,...,p_n,\lambda_n\right) \right|^2
\eq
with $\lambda_i \in \{+,-\}$.
The amplitudes ${\cal A}^{(0)}_n\left(p_1,\lambda_1,...,p_n,\lambda_n\right)$ are called helicity amplitudes.
The spin-summed squared matrix element $| {\cal A}^{(0)}_n |^2$
involves therefore the evaluation of $2^n$ helicity amplitudes.
For multi-parton amplitudes it is desirable to avoid the exponential growth due to the spin sum.
One possibility is to replace the spin-summed squared matrix element by
a random sampling of squared helicity amplitudes. In combination with the phase space integral, which usually
is done by Monte Carlo techniques, one would therefore write
\bq
\lefteqn{
 \int\limits_n d\phi_n {\cal O}_n  \left| {\cal A}^{(0)}_{n+2}\left(p_1,...,p_{n+2} \right) \right|^2
 = } & &
 \\
 & &
 \int\limits_{[0,1]^{n+2}} d^{n+2}u
 \int\limits_n d\phi_n {\cal O}_n 
 \left| {\cal A}^{(0)}_{n+2}\left(p_1,\lambda(u_1),...,p_{n+2},\lambda(u_{n+2})\right) \right|^2,
 \nonumber
\eq
with
\bq
 \lambda(u) & = & \left\{ \begin{array}{lll} + & \mbox{for} & 0 \le u < \frac{1}{2}, \\
                                            - & \mbox{for} & \frac{1}{2} \le u < 1. \\
                          \end{array} \right.
\eq
We refer to this method as ``helicity sampling''.
Within this method, the integrand as a function of the variables $u_1$, ..., $u_{n+2}$ 
is discontinuous at $u_i=1/2$.

A second possibility avoids this discontinuity and is referred to as the method of ``random polarisations''.
We discuss this method first for gluons: Instead of polarisation vectors $\eps_\mu^+$ and $\eps_\mu^-$
with definite helicity ``$+$'' and ``$-$'' one introduces linear combinations \cite{Draggiotis:1998gr}
\bq
\label{def_random_polarisation}
 \eps_\mu(\phi) & = & e^{i\phi} \eps_\mu^+ \; + \; e^{-i\phi} \eps_\mu^-.
\eq
We may then replace the summation over the spin states by an integration over the angle $\phi$.
\bq
 \sum\limits_{\lambda=\pm} {\eps_\mu^\lambda}^\ast \eps_\nu^\lambda 
 & = & 
 \frac{1}{2\pi} \int\limits_0^{2\pi} d\phi \; {\eps_\mu(\phi)}^\ast \eps_\nu(\phi)
\eq
We thus arrive at
\bq
 \int\limits_n d\phi_n {\cal O}_n  \left| {\cal A}^{(0)}_{n+2}\left(p_1,...,p_{n+2} \right) \right|^2
 & = & 
 \int\limits_{[0,1]^{n+2}} d^{n+2}u
 \int\limits_n d\phi_n {\cal O}_n 
 \left| {\cal A}^{(0)}_{n+2}\left(p_1,\phi_1,...,p_{n+2},\phi_{n+2}\right) \right|^2,
 \nonumber \\
\eq
with $\phi_i= 2 \pi u_i$. The integrand is now a smooth function of the additional variables $u_i$.
The method is easily extended to quarks. One defines the linear combinations
\bq
 u(\phi) = e^{-i\phi} u^+ \; + \; e^{i \phi} u^-,
 & &
 \bar{u}(\phi) = e^{i\phi} \bar{u}^+ \; + \; e^{-i \phi} \bar{u}^-,
 \nonumber \\
 v(\phi) = e^{-i\phi} v^+ \; + \; e^{i \phi} v^-,
 & &
 \bar{v}(\phi) = e^{i\phi} \bar{v}^+ \; + \; e^{-i \phi} \bar{v}^-.
\eq
Then
\bq
 \sum\limits_{\lambda=\pm} u^\lambda \bar{u}^\lambda 
 = \frac{1}{2\pi} \int\limits_0^{2\pi} d\phi \; u(\phi) \bar{u}(\phi),
 & &
 \sum\limits_{\lambda=\pm} v^\lambda \bar{v}^\lambda 
 = \frac{1}{2\pi} \int\limits_0^{2\pi} d\phi \; v(\phi) \bar{v}(\phi).
\eq
We have outlined the method of random polarisations for the Born contribution.
The method of random polarisations works without modification for the virtual contribution and the insertion term as well.
However, it will not work without modifications for the subtracted real contribution.
The reason is the following: The dipole subtraction terms are constructed by requiring to match locally the singular behaviour of the 
spin-summed squared matrix element of the real contribution.
If we consider instead individual random polarisations, there is no guarantee that the subtraction terms match locally the singular behaviour,
in general they will do so only after integration over the additional angles.
Therefore, if one wants to use the method of random polarisations for the subtracted real contribution, one has to add additional subtraction
terms which ensure that the integrand is locally integrable.
We note that in ref.~\cite{Czakon:2009ss,Dittmaier:2008md} subtraction terms for helicity eigenstates have been derived. These subtraction terms allow to use the method
of helicity sampling for the real contribution, but not the method of random polarisations.
This can be seen as follows: For example, for the gluon polarisation vectors one has
\bq
\label{random_pol_gluon}
 \eps_\mu(\phi)^\ast \eps_\nu(\phi) 
 & = &
 \left[ e^{-i\phi} {\eps_\mu^+}^\ast + e^{i\phi} {\eps_\mu^-}^\ast \right] \left[ e^{i\phi} \eps_\nu^+ + e^{-i\phi} \eps_\nu^- \right]
 \nonumber \\
 & = &
 {\eps_\mu^+}^\ast \eps_\nu^+
 +
 {\eps_\mu^-}^\ast \eps_\nu^-
 +
 e^{-2i\phi} {\eps_\mu^+}^\ast \eps_\nu^-
 +
 e^{2i\phi} {\eps_\mu^-}^\ast \eps_\nu^+.
\eq
Ref.~\cite{Czakon:2009ss,Dittmaier:2008md} provides local subtraction terms for the contributions proportional to 
${\eps_\mu^+}^\ast \eps_\nu^+$ and ${\eps_\mu^-}^\ast \eps_\nu^-$,
but not for the ones proportional to the mixed combinations ${\eps_\mu^+}^\ast \eps_\nu^-$ and ${\eps_\mu^-}^\ast \eps_\nu^+$.
In this paper we derive the local subtraction terms required to combine the dipole subtraction scheme with the method of random polarisations.
We note that in eq.~(\ref{random_pol_gluon}) the terms
\bq
 {\eps_\mu^+}^\ast \eps_\nu^+
 +
 {\eps_\mu^-}^\ast \eps_\nu^-
\eq
on the right-hand side are just the usual spin sum.
These are approximated by the spin-summed dipole subtraction terms.
In combining the dipole subtraction scheme with the method of random polarisations we need in addition subtraction terms corresponding 
to the terms
\bq
 e^{-2i\phi} {\eps_\mu^+}^\ast \eps_\nu^-
 & \mbox{and} & 
 e^{2i\phi} {\eps_\mu^-}^\ast \eps_\nu^+.
\eq
In an individual dipole subtraction term corresponding to the splitting $(ij) \rightarrow i +j$ with spectator $k$ we will have to include the dependence
on the helicity angles $\phi_i$ and $\phi_j$ of the particles $i$ and $j$. The helicity angle $\phi_k$ of the spectator does not enter
the dipole subtraction terms.
The terms independent of $\phi_i$ and $\phi_j$ will just be the spin-summed dipole subtraction terms, which are well-known 
and given in ref.~\cite{Catani:1997vz,Catani:1997vzerr,Catani:2002hc}.
Therefore we only have to derive the dipole subtraction terms with a non-trivial dependence on $\phi_i$ and $\phi_j$.
Since the dependence on $\phi_i$ and $\phi_j$ is only through the exponential factors
\bq
 e^{\pm 2 i \phi_i},
 & &
 e^{\pm 2 i \phi_j},
\eq
it is clear that the new subtraction terms integrate to zero after the integration over
the helicity angles $\phi_i$ and $\phi_j$.
We therefore modify only the subtraction terms for the real part, but not the integrated subtraction terms.
We also note that for the method of random polarisations the local subtraction terms for the individual helicity eigenstates
$ {\eps_\mu^+}^\ast \eps_\nu^+$ and ${\eps_\mu^-}^\ast \eps_\nu^-$
need not be known.

We would like to add a comment on dimensional regularisation.
In the discussion above we have assumed for all particles two helicity eigenstates, as if we would work in four dimensions.
This is justified by the following argument:
The subtraction terms are required to match the singular behaviour of the integrand in $D$ dimensions.
Throughout this paper we use $D=4-2\eps$.
There are several variants of dimensional regularisation. All of them continue the integration over the loop momenta and the integration
over the phase space of unobserved particles to $D$ dimensions. 
They differ in how they treat the spin degrees of freedom of the particles (and the momenta of the observed particles).
Conventional dimensional regularisation (CDR) continues all spin degrees of freedom to $D$ dimensions, while the
't Hooft-Veltman scheme (HV) continues only the spin degrees of freedom of the unobserved particles to $D$ dimensions and keeps the ones of
the observed particles in $4$ dimensions.
The four-dimensional helicity scheme (FDH) keeps all spin degrees of freedom in $4$ dimensions.
In the CDR scheme and in the HV scheme the unobserved particles may have additional spin degrees of freedom, which are referred to as
$\eps$-helicities.
We will discuss the case of a gluon.
Let us denote the polarisation vectors corresponding to the additional $\eps$-helicities by
\bq
 \eps_\mu^{(-2\eps)}
\eq
The spin sum is then given by
\bq
 {\eps_\mu^+}^\ast \eps_\nu^+
 +
 {\eps_\mu^-}^\ast \eps_\nu^-
 +
 {\eps_\mu^{(-2\eps)}}^\ast \eps_\nu^{(-2\eps)}.
\eq
In the four-dimensional sub-space we may replace the sum over the helicities ``$+$'' and ``$-$'' by an integration over the random polarisation
$\eps_\mu(\phi)$.
We thus have
\bq
 \eps_\mu(\phi)^\ast \eps_\nu(\phi) 
 +
 {\eps_\mu^{(-2\eps)}}^\ast \eps_\nu^{(-2\eps)}
 & = &
 {\eps_\mu^+}^\ast \eps_\nu^+
 +
 {\eps_\mu^-}^\ast \eps_\nu^-
 +
 {\eps_\mu^{(-2\eps)}}^\ast \eps_\nu^{(-2\eps)}
 +
 e^{-2i\phi} {\eps_\mu^+}^\ast \eps_\nu^-
 +
 e^{2i\phi} {\eps_\mu^-}^\ast \eps_\nu^+.
 \nonumber 
\eq
The right-hand side is again given as the ($D$-dimensional) spin sum plus terms proportional to ${\eps_\mu^+}^\ast \eps_\nu^-$ and
${\eps_\mu^-}^\ast \eps_\nu^+$.
By construction the subtracted real contribution is integrable in four dimensions.
Therefore we can take the limit $D \rightarrow 4$. In taking this limit all terms proportional to $\eps_\mu^{(-2\eps)}$ disappear.
In summary, we have seen that for the extension of the dipole subtraction method to the method of random polarisations it is sufficient
to consider the four-dimensional helicities ``$+$'' and ``$-$'' only.
In the discussion above we have used the gluon polarisations as an example.
The discussion for the polarisation states of the quarks and the anti-quarks is similar.

\section{Soft and collinear limits}
\label{sec:limits}

In this section we discuss the behaviour of $|{\cal A}_{n+1}^{(0)}|^2$ in the soft and collinear limits.
In the soft limit we parametrise the momentum of the soft parton $p_j$ as
\bq
 p_j & = & \lambda q
\eq
and consider contributions to $|{\cal A}_{n+1}^{(0)}|^2$ of the order $\lambda^{-2}$.
Contributions to $|{\cal A}_{n+1}^{(0)}|^2$ which are less singular than $\lambda^{-2}$ are integrable in the soft limit.
In the collinear $p_i || p_j$ limit we parametrise the momenta of two collinear massless final-state partons $i$ and $j$ as
\bq
\label{collinearlimit}
p_i & = & z p + k_\perp - \frac{k_\perp^2}{z} \frac{n}{2 p \cdot n }, \nonumber \\
p_j & = & (1-z)  p - k_\perp - \frac{k_\perp^2}{1-z} \frac{n}{2 p \cdot n }.
\eq
Here $n$ is a massless four-vector and the transverse component $k_\perp$ satisfies
$2pk_\perp = 2n k_\perp =0$.
The four-vectors $p$, $p_i$ and $p_j$ are massless: $p^2=p_i^2=p_j^2=0$.
The collinear limit occurs for $k_\perp^2 \rightarrow 0$.
We consider contributions to $|{\cal A}_{n+1}^{(0)}|^2$ of the order $|k_\perp|^{-2}$.
Contributions to $|{\cal A}_{n+1}^{(0)}|^2$ which are less singular than $|k_\perp|^{-2}$ are integrable in the collinear limit.

The collinear limit, which occurs in massless theories has a generalisation to massive theories, called the
quasi-collinear limit. 
In the quasi-collinear limit we consider a splitting $\tilde{ij} \rightarrow i + j$, with associated 
particle masses $m_{ij}$, $m_i$ and $m_j$.
In the quasi-collinear limit we parametrise the momenta as
\bq
 p_i & = & z p + k_\perp - \frac{k_\perp^2+z^2 m_{ij}^2 - m_i^2}{z} \frac{n}{2 p \cdot n},
 \nonumber \\
 p_j & = & (1-z)  p - k_\perp - \frac{k_\perp^2+(1-z)^2m_{ij}^2-m_j^2}{1-z} \frac{n}{2 p \cdot n }.
\eq
Again, $n$ is a massless four-vector and the transverse component $k_\perp$ satisfies
$2pk_\perp = 2n k_\perp =0$.
The four-vectors $p$, $p_i$ and $p_j$ are on-shell:
\bq
 p^2 = m_{ij}^2,
 \;\;\;\;\;\;
 p_i^2 = m_i^2,
 \;\;\;\;\;\;
 p_j^2 = m_j^2.
\eq
In the quasi-collinear limit we take terms of the order ${\cal O}(k_\perp)$, 
${\cal O}(m_{ij})$, ${\cal O}(m_i)$ and ${\cal O}(m_j)$
to be of the same order
\bq
 {\cal O}(k_\perp)
 & = & 
 {\cal O}(m_{ij})
 =
 {\cal O}(m_{i})
 =
 {\cal O}(m_{j})
\eq
and consider terms which correspond to the order of $|k_\perp|^{-2}$.
Obviously, the collinear limit is a special case of the quasi-collinear limit.
\\
\\
If the emitting particle is in the initial state, the collinear limit is defined as
\bq
 p_a & = & p,
 \nonumber \\
 p_i & = & \left(1-x\right) p + k_\perp - \frac{k_\perp^2}{1-x} \frac{n}{2 p \cdot n},
 \nonumber \\
 p_{a i} & = & x p - k_\perp - \frac{k_\perp^2}{x} \frac{n}{2 p \cdot n}.
\eq
Here, all particles are massless. In this paper we restrict ourselves to massless incoming partons, therefore
we do not have to consider the generalisation to the massive quasi-collinear case for initial-state partons.
\\
\\
In the soft limit a Born amplitude ${\cal A}_{n+1}^{(0)}$ with $(n+1)$ partons behaves as
\bq
 \lim\limits_{p_j \rightarrow 0} {\cal A}_{n+1}^{(0)}
 & = & 
 g \mu^\eps \eps_\mu(p_j) {\bf J}^\mu
 {\cal A}_{n}^{(0)}.
\eq
Here, $g$ denotes the strong coupling, $\mu$ is a scale introduced to keep the coupling dimensionless, $p_j$ is the momentum of the soft gluon.
The eikonal current is given by
\bq
 {\bf J}^\mu & = & \sum\limits_{i=1}^n {\bf T}_i \frac{p_i^\mu}{p_i \cdot p_j}.
\eq
The sum is over the remaining $n$ hard momenta $p_i$.
The colour charge operators ${\bf T}_i$ for the emission of a gluon from a quark, gluon or antiquark in the final state are defined by
\bq
\label{colour_charge_operator_final}
\mbox{quark :} & & 
 {\bf T}_{q \rightarrow q g} {\cal A}\left(  ... q_j ... \right) =
 \left( T_{ij}^a \right) {\cal A}\left(  ... q_j ... \right), \nonumber \\
\mbox{gluon :} & & 
 {\bf T}_{g \rightarrow g g}{\cal A}\left(  ... g^b ... \right) =
 \left( i f^{cab} \right) {\cal A}\left(  ... g^b ... \right), \nonumber \\
\mbox{antiquark :} & & 
 {\bf T}_{\bar{q} \rightarrow \bar{q} g} {\cal A}\left(  ... \bar{q}_j ... \right) =
 \left( - T_{ji}^a \right) {\cal A}\left(  ... \bar{q}_j ... \right).
\eq
The minus sign for the antiquark has its origin in the fact that for an outgoing antiquark the (outgoing) 
momentum flow is opposite to the flow of the fermion line.
The corresponding colour charge operators for the emission of a gluon from a quark, gluon or antiquark in the initial state are
\bq
\label{colour_charge_operator_initial}
\mbox{quark :} & & 
 {\bf T}_{\bar{q} \rightarrow \bar{q} g} {\cal A}\left(  ... \bar{q}_j ... \right) =
 \left( - T_{ji}^a \right) {\cal A}\left(  ... \bar{q}_j ... \right), \nonumber \\
\mbox{gluon :} & & 
 {\bf T}_{g \rightarrow g g} {\cal A}\left(  ... g^b ... \right) =
 \left( i f^{cab} \right) {\cal A}\left(  ... g^b ... \right), \nonumber \\
\mbox{antiquark :} & & 
 {\bf T}_{q \rightarrow q g}{\cal A}\left(  ... q_j ... \right) =
 \left( T_{ij}^a \right) {\cal A}\left(  ... q_j ... \right). 
\eq
In the amplitude an incoming quark is denoted as an outgoing antiquark and vice versa.
For the squares of the colour charge operators one has
\bq
 {\bf T}_{q \rightarrow q g}^2 = C_F,
 & &
 {\bf T}_{g \rightarrow g g}^2 = C_A.
\eq
We also define the colour charge operator for the emission of a quark-antiquark pair from a gluon by
\bq
 {\bf T}_{g \rightarrow q \bar{q}} {\cal A}\left(  ... g^b ... \right) 
 & = &
 \left( T^b_{ij} \right) {\cal A}\left(  ... g^b ... \right)
\eq
and
\bq
 {\bf T}_{g \rightarrow q \bar{q}}^2 = T_R.
\eq
$C_A$, $C_F$ and $T_R$ are the usual $SU(N_c)$ colour factors
\bq
 C_A = N_c,
 \;\;\;\;\;\;
 C_F = \frac{N_c^2-1}{2 N_c},
 \;\;\;\;\;\;
 T_R = \frac{1}{2}.
\eq 
In squaring an amplitude we obtain terms proportional to ${\bf T}_i \cdot {\bf T}_k$ ($k \neq i$)
and terms proportional to ${\bf T}_i^2$.
In order to obtain the dipole structure we would like to re-express ${\bf T}_i^2$ as a combination of
terms involving only ${\bf T}_i \cdot {\bf T}_k$ with $k \neq i$.
This can be done using colour conservation. We write for $i \in \{q, g\}$
\bq
 {\bf T}_i^2 & = & 
 - \sum\limits_{k=1, k \neq i}^n
 {\bf T}_i \cdot {\bf T}_k.
\eq
For the splitting $g \rightarrow q \bar{q}$ we write
\bq
 {\bf T}_{g \rightarrow q \bar{q}}^2 & = & 
 - \sum\limits_{k=1, k \neq i}^n
 \frac{{\bf T}_{g \rightarrow q \bar{q}}^2}{{\bf T}_{i}^2}
 {\bf T}_i \cdot {\bf T}_k.
\eq
In the following we will often encounter functions which depend on the polarisation of a particle, for example in the case of a gluon
\bq
 f\left(\eps_\mu \right).
\eq
If the polarisation vector $\eps_\mu$ corresponds to a random polarisation with helicity angle $\phi$ according to eq.~(\ref{def_random_polarisation})
then we simply write $f\left(\phi\right)$ instead of $f\left(\eps_\mu\left(\phi\right) \right)$, i.e.
\bq
 f\left(\phi\right) & = &  
 f\left(\eps_\mu\left(\phi\right) \right).
\eq
If on the other hand $\eps_\mu$ corresponds to a helicity eigenstate $\eps_\mu^\lambda$ we write
\bq
 f\left(\lambda\right)
  & = &
 f\left(\eps_\mu^\lambda\right).
\eq
We use the notation
\bq
 f\left(h\right)
\eq
to denote either $f\left(\phi\right)$ or $f\left(\lambda\right)$.
This allows for a uniform notation for the polarisations of quarks and gluons.

Squaring the soft limit of the amplitude and using colour conservation one finds
\bq
 \lim\limits_{p_j \rightarrow 0} \left| {\cal A}_{n+1}^{(0)} \right|^2
 & = &
 - 4 \pi \alpha_s \mu^{2\eps}
 \; \sum\limits_{i=1}^n \;
 \sum\limits_{\substack{k=1\\ k \neq i}}^n \;
 {{\cal A}_{n}^{(0)}}^\ast
 {\bf T}_i \cdot {\bf T}_k
 S_{ij,k}\left(\eps_j\right)
 {\cal A}_{n}^{(0)},
\eq
with
\bq
 S_{ij,k}\left(\eps_j\right)
 & = & 
 \frac{\left( p_i \cdot \eps_j^\ast \right) \left( p_i\cdot \eps_j \right)}{\left(p_i \cdot p_j \right)^2}
 -
 \frac{\left( p_i \cdot \eps_j^\ast \right) \left( p_k \cdot \eps_j \right) + \left( p_k \cdot \eps_j^\ast \right) \left(p_i \cdot \eps_j \right)}
      { \left( p_i \cdot p_j \right) \left( p_i \cdot p_j + p_j \cdot p_k\right)}.
\eq
We will need the (quasi)-collinear limit of the singular soft function $S_{ij,k}\left(\eps_j\right)$. We find
\bq
 \lim\limits_{p_i || p_j}
 S_{ij,k}\left(\eps_j\right) & = &
 \frac{\left( p_i \cdot \eps_j^\ast \right) \left( p_i\cdot \eps_j \right)}{\left(p_i \cdot p_j \right)^2}.
\eq
The second term of $S_{ij,k}\left(\eps_j\right)$ scales like $|k_\perp|^{-1}$ and is therefore not singular enough in the (quasi)-collinear limit.
Within the method of random polarisations we are in particular interested in the terms with a non-trivial dependence on $\phi_j$.
From eq.~(\ref{random_pol_gluon}) it follows that these terms are given by
\bq
 S_{ij,k}\left(\phi_j\right)
 - 
 \sum\limits_{\lambda_j} S_{ij,k}\left(\lambda_j\right).
\eq
Explicitly, we have
\bq
\lefteqn{
 S_{ij,k}\left(\phi_j\right)
 - 
 \sum\limits_{\lambda_j} S_{ij,k}\left(\lambda_j\right)
 = } & & 
 \\
 & &
 e^{-2i\phi_j}
 \left[
 \frac{\left( p_i \cdot {\eps_j^+}^\ast \right) \left( p_i\cdot \eps_j^- \right)}{\left(p_i \cdot p_j \right)^2}
 -
 \frac{\left( p_i \cdot {\eps_j^+}^\ast \right) \left( p_k \cdot \eps_j^- \right) + \left( p_k \cdot {\eps_j^+}^\ast \right) \left(p_i \cdot \eps_j^- \right)}
      { \left( p_i \cdot p_j \right) \left( p_i \cdot p_j + p_j \cdot p_k\right)}
 \right] 
 \nonumber \\
 & &
 +
 e^{2i\phi_j}
 \left[
 \frac{\left( p_i \cdot {\eps_j^-}^\ast \right) \left( p_i\cdot \eps_j^+ \right)}{\left(p_i \cdot p_j \right)^2}
 -
 \frac{\left( p_i \cdot {\eps_j^-}^\ast \right) \left( p_k \cdot \eps_j^+ \right) + \left( p_k \cdot {\eps_j^-}^\ast \right) \left(p_i \cdot \eps_j^+ \right)}
      { \left( p_i \cdot p_j \right) \left( p_i \cdot p_j + p_j \cdot p_k\right)}
 \right].
 \nonumber
\eq
Let us now turn to the quasi-collinear limit. 
We discuss immediately the quasi-collinear limit, the collinear limit being a special case of the quasi-collinear limit.
It will be sufficient to discuss the case where all particles are in the final state.
The collinear limit in the initial state case can be obtained from the collinear limit in the final state by crossing.
In the quasi-collinear limit the Born amplitude factorises according to
\bq
\label{collinear_factorisation_amplitude}
\lefteqn{
 \lim\limits_{p_i || p_j}
 {\cal A}_{n+1}^{(0)}\left(...,p_i,...,p_j,...\right) 
 = } & &
 \nonumber \\
 & & 
 g \mu^\eps
 \sum\limits_{\lambda} \; \mbox{Split}_{(ij) \rightarrow i+j}^\lambda(p_{ij},p_i,p_j,h_i,h_j) \; {\bf T}_{(ij) \rightarrow i+j} \; {\cal A}_{n}^{(0)}\left(...,p_{ij},\lambda,...\right).
\eq
where the sum is over all polarisations of the intermediate particle.
The splitting amplitudes $\mbox{Split}$ are universal, they depend
only on the two momenta becoming (quasi)-collinear, and not upon the specific amplitude under
consideration.
The splitting functions $\mbox{Split}$ are given by
\bq
\label{def_split}
%
\mbox{Split}^\lambda_{q \rightarrow q g}\left(p,p_i,p_j,h_i,h_j\right) & = &
 \frac{1}{(p_i + p_j)^2 - m_{ij}^2} \bar{u}(p_i) \eps\!\!\!/(p_j) u^\lambda(p),
\nonumber \\
\mbox{Split}^\lambda_{g \rightarrow g g}\left(p,p_i,p_j,h_i,h_j\right) & = &
 \frac{2}{2 p_i \cdot p_j} \left[
    \eps(p_i) \cdot \eps(p_j) \; p_i \cdot \left. \eps^\lambda(p) \right.^\ast
  + \eps(p_j) \cdot \left. \eps^\lambda(p) \right.^\ast \; p_j \cdot \eps(p_i)
 \right. \nonumber \\
 & & \left.
  - \eps(p_i) \cdot \left. \eps^\lambda(p) \right.^\ast \; p_i \cdot \eps(p_j)
 \right],
\nonumber \\
\mbox{Split}^\lambda_{g \rightarrow q \bar{q}}\left(p,p_i,p_j,h_i,h_j\right) & = &
 \frac{1}{2 p_i \cdot p_j} \bar{u}(p_i) \left. \eps\!\!\!/^\lambda(p) \right.^\ast v(p_j).
\eq
Here we used the notation $\left. \eps\!\!\!/^\lambda(p) \right.^\ast = \eps_\mu^\lambda(p)^\ast \; \gamma^\mu$, i.e. complex conjugation is only with respect to the polarisation vector.
These formulae hold for any polarisation of the particles $i$, $j$ and $(ij)$.
We will use these formulae where particles $i$ and $j$ have random polarisations, while the mother particle $(ij)$ 
has a definite polarisation $\lambda = +$ or $\lambda = -$.
There is no point in introducing a random polarisation for the mother particle $(ij)$, since eq.~(\ref{collinear_factorisation_amplitude})
involves a sum over the intermediate polarisation (or equivalently an integration of the intermediate polarisation).
We define the squares of the splitting amplitudes by
\bq
\label{def_P}
 \left[ P_{(ij) \rightarrow i+j}\left(p,p_i,p_j,h_i,h_j\right) \right]_{\alpha\beta} & = &  
 \sum\limits_{\lambda,\lambda'}
 u_\alpha^\lambda(p) \left. \; \mbox{Split}^\lambda \right.^\ast \mbox{Split}^{\lambda'} \;\bar{u}_\beta^{\lambda'}(p)
 \;\;\;\;\mbox{for quarks,}
 \nonumber \\
 \left[ P_{(ij) \rightarrow i+j}\left(p,p_i,p_j,h_i,h_j\right)\right]_{\mu\nu} & = &  
 \sum\limits_{\lambda,\lambda'}
 \left. \eps_\mu^\lambda(p) \right.^\ast \;
                \left. \mbox{Split}^\lambda \right.^\ast \mbox{Split}^{\lambda'} \; \eps_\nu^{\lambda'}(p)
 \;\;\;\;\mbox{for gluons.}
\eq
Let us denote by ${\cal A}_{n}^{\xi \;(0)}$ the amplitude, where the polarisation vector of particle $(ij)$ has been removed.
If particle $(ij)$ is a gluon, $\xi$ is a Lorentz index, while in the case where particle $(ij)$ is a quark $\xi$ corresponds to a Dirac index.
With this notation the squared amplitude factorises in the collinear limit as
\bq
\label{amplitude_squared_coll}
 \lim\limits_{p_i || p_j}
 \left|{\cal A}_{n+1}^{(0)} \right|^2 & = & 
 4 \pi \alpha_s \mu^{2\eps} 
 {{\cal A}_{n}^{\xi \;(0)}}^\ast
 {\bf T}_{(ij) \rightarrow i+j}^2
 \left[ P_{(ij) \rightarrow i+j}\left(p,p_i,p_j,h_i,h_j\right) \right]_{\xi \xi'}
 {\cal A}_{n}^{\xi' \;(0)}.
\eq
Again we are in particular interested in the terms with a non-trivial dependence on $\phi_i$ or $\phi_j$.
These terms are given by
\bq
\label{collinear_splitting_non_trivial_phi}
 P_{(ij) \rightarrow i+j}\left(\phi_i,\phi_j\right)
 -
 \sum\limits_{\lambda_i, \lambda_j}
 P_{(ij) \rightarrow i+j}\left(\lambda_i,\lambda_j\right).
\eq
The second term in eq.~(\ref{collinear_splitting_non_trivial_phi}) 
corresponds to the splitting functions summed over the polarisations of the collinear particles $i$ and $j$.
Subtracting out the sum over the polarisations of two particles is an operation which will occur more often in this paper and it is
convenient to introduce a short-hand notation for that. We define the operator ${\cal R}$ acting on a function $f(h_i,h_j)$ depending on two polarisations by
\bq
\label{def_R_operation}
 {\cal R} f\left(\phi_i,\phi_j\right)
 & = & 
 f \left(\phi_i,\phi_j\right)
 - 
 \sum\limits_{\lambda_i, \lambda_j}
 f\left(\lambda_i,\lambda_j\right).
\eq
Let us for a moment specialise to the massless case.
For the singular part of the splitting functions one has then the well-known expressions
\bq
\label{P_spin_summed}
 \sum\limits_{\lambda_i, \lambda_j}
 P_{q \rightarrow q g }\left(\lambda_i,\lambda_j\right) & = & 
   \frac{2}{2 p_i \cdot p_j} p\!\!\!/ \left[ \frac{2z}{1-z} + (1 - \eps) (1-z) \right]
 + {\cal O}\left(\frac{1}{\left|k_\perp\right|}\right), \nonumber \\
 \sum\limits_{\lambda_i, \lambda_j}
 P_{g \rightarrow g g}\left(\lambda_i,\lambda_j\right) & = & 
   \frac{2}{2 p_i \cdot p_j}  \left[ - g^{\mu\nu} \left( \frac{2z}{1-z} + \frac{2(1-z)}{z} \right) 
   - 4 (1-\eps) z (1-z) \frac{k^\mu_\perp k^\nu_\perp}{k_\perp^2} \right]
 \nonumber \\
 & &
+ {\cal O}\left(\frac{1}{\left|k_\perp\right|}\right), \nonumber \\
 \sum\limits_{\lambda_i, \lambda_j}
 P_{g \rightarrow q \bar{q}}\left(\lambda_i,\lambda_j\right) & = & 
   \frac{2}{2 p_i \cdot p_j} \left[ -g^{\mu\nu} + 4 z (1-z) \frac{k^\mu_\perp k^\nu_\perp}{k_\perp^2} \right]
+ {\cal O}\left(\frac{1}{\left|k_\perp\right|}\right).
\eq
Note that these equations hold up to terms of order $1/|k_\perp|$. Terms of order $1/|k_\perp|$ are integrable in the collinear limit.
We have defined the splitting functions through eq.~(\ref{def_split}) and eq.~(\ref{def_P}).
Through these definitions we pick up integrable terms not shown explicitly on the r.h.s of eq.~(\ref{P_spin_summed}).

The quasi-collinear splittings $q \rightarrow q g$ and $ g \rightarrow g g$ have non-vanishing soft limits.
We consider the case where particle $j$ becomes soft.
We find
\bq
 \lim\limits_{p_j \rightarrow 0}  \left[ P_{q \rightarrow q g} \right]_{\alpha \beta}
 & = & 
 \frac{\left( p_i \cdot \eps_j^\ast \right) \left( p_i\cdot \eps_j \right)}{\left(p_i \cdot p_j \right)^2}
 \;
 u_\alpha(p_i) \bar{u}_\beta(p_i),
 \nonumber \\
 \lim\limits_{p_j \rightarrow 0}  \left[ P_{g \rightarrow g g} \right]_{\mu\nu}
 & = & 
 \frac{\left( p_i \cdot \eps_j^\ast \right) \left( p_i\cdot \eps_j \right)}{\left(p_i \cdot p_j \right)^2}
 \;
 \left. \eps_\mu(p_i) \right.^\ast \eps_\nu(p_i).
\eq
This shows that the (quasi)-collinear limit of the soft singular function agrees with the soft limit of the (quasi)-collinear singular function.
To construct the subtraction terms we can therefore start from the (quasi)-collinear singular function 
and supplement it with the terms which are singular in the soft limit,
but not in the (quasi)-collinear limit.
It is convenient to define two functions for the terms singular in the soft limit, but not in the (quasi)-collinear limit.
We set
\bq
\label{def_S}
 \left[ S_{q \rightarrow q g}\left(p,p_i,p_j,p_k,h_i,h_j\right) \right]_{\alpha \beta} & = &  
 -
 \frac{\left( p_i \cdot \eps_j^\ast \right) \left( p_k \cdot \eps_j \right) + \left( p_k \cdot \eps_j^\ast \right) \left(p_i \cdot \eps_j \right)}
      { \left( p_i \cdot p_j \right) \left( p_i \cdot p_j + p_j \cdot p_k\right)}
 \;
 u_\alpha(p_i) \bar{u}_\beta(p_i),
 \nonumber \\
 \left[ S_{g \rightarrow g g}\left(p,p_i,p_j,p_k,h_i,h_j\right) \right]_{\mu\nu} & = &  
 -
 \frac{\left( p_i \cdot \eps_j^\ast \right) \left( p_k \cdot \eps_j \right) + \left( p_k \cdot \eps_j^\ast \right) \left(p_i \cdot \eps_j \right)}
      { \left( p_i \cdot p_j \right) \left( p_i \cdot p_j + p_j \cdot p_k\right)}
 \;
 \left. \eps_\mu(p_i) \right.^\ast \eps_\nu(p_i).
 \nonumber \\
 & &
 -
 \frac{\left( p_j \cdot \eps_i^\ast \right) \left( p_k \cdot \eps_i \right) + \left( p_k \cdot \eps_i^\ast \right) \left(p_j \cdot \eps_i \right)}
      { \left( p_i \cdot p_j \right) \left( p_i \cdot p_j + p_i \cdot p_k\right)}
 \;
 \left. \eps_\mu(p_j) \right.^\ast \eps_\nu(p_j).
 \nonumber \\
\eq

\section{The subtraction terms}
\label{sec:subtraction_terms}

In this section we give the additional subtraction terms, which extend the dipole subtraction method 
to random polarisations of the external partons.
We recall that the NLO-contribution to an observable in the spin-summed case is given by
\bq
\label{standard_NLO_subtraction}
\langle O \rangle^{NLO} & = & 
 \int\limits_{n+1} \left( O_{n+1} d\sigma^R - O_n d\sigma^A \right)
 + \int\limits_n \left( O_n d\sigma^V + O_n d\sigma^C + O_n \int\limits_1 d\sigma^A \right).
\eq
We modify this scheme by including an additional subtraction term $d\sigma^{\tilde{A}}$
\bq
\label{modified_NLO_subtraction}
\langle O \rangle^{NLO} & = & 
 \int\limits_{n+1} \left[ O_{n+1} d\sigma^R - O_n \left( d\sigma^A + d\sigma^{\tilde{A}} \right) \right]
 + \int\limits_n \left( O_n d\sigma^V + O_n d\sigma^C + O_n \int\limits_1 d\sigma^A \right).
 \;\;\;\;
\eq
The additional subtraction $d\sigma^{\tilde{A}}$ term ensures 
that the expression in the square bracket in the first term of eq.~(\ref{modified_NLO_subtraction}) is locally integrable
when used with random polarisations.
$d\sigma^{\tilde{A}}$ is chosen such that
\bq
 \frac{1}{\left(2\pi\right)^2} \int\limits_0^{2\pi} d\phi_i \int\limits_0^{2\pi}d\phi_j d\sigma^{\tilde{A}}
 & = & 0,
\eq
where $\phi_i$ and $\phi_j$ are the helicity angles of the two unresolved particles.
Therefore $d\sigma^{\tilde{A}}$ is a function, which integrates to zero and can be added to eq.~(\ref{standard_NLO_subtraction}).
Similar to the standard case, $d\sigma^{\tilde{A}}$ is given as a sum of dipoles:
\bq
 d\sigma^{\tilde{A}}
 & = &
 \sum\limits_{(i,j)} \sum\limits_{k\neq i,j} \tilde{\cal D}_{ij,k}
 +
 \sum\limits_{(i,j)} \sum\limits_{a} \tilde{\cal D}_{ij}^a
 +
 \sum\limits_{(a.j)} \sum\limits_{k\neq j} \tilde{\cal D}^{aj}_{k}
 +
 \sum\limits_{(a,j)} \sum\limits_{b\neq a} \tilde{\cal D}^{aj,b}.
\eq
As in the spin-summed case, the dipoles consist of dipole splitting functions sandwiched between
tree-level amplitudes with $n$ partons.
We can use crossing symmetry  to obtain all additional dipole splitting functions from the final-final
case.
(The tree-level $n$-parton amplitudes are evaluated with mapped momenta, and the actual form
of this mapping depends on whether the partons are in the initial or final state.)
In connection with crossing symmetry it is useful to define the following operation
\bq
\label{def_C_operation}
 {\cal C}_i & : & 
 \eps_i \leftrightarrow \eps^\ast_i,
 \nonumber \\
 & & \bar{u}_i \leftrightarrow \bar{v}_i,
 \nonumber \\
 & & u_i \leftrightarrow v_i,
\eq
which adjusts the polarisation vector or spinor of the $i$-th particle from the final to the initial state and vice versa.

\subsection{The additional subtraction terms}

\subsubsection{Final-state emitter and final-state spectator}

If both the emitter and the spectator are in the final state, the additional subtraction terms are given by
\bq
\lefteqn{
\tilde{\cal D}_{ij,k}
 = 
 - 4 \pi \alpha_s \mu^{2\eps} 
 } & & 
 \\
 & &
 {{\cal A}_{n}^{\xi \;(0)}}\left(...,\tilde{p}_{ij},...,\tilde{p}_{k},...\right)^\ast
 \;\;
 \frac{{\bf T}_{ij} \cdot {\bf T}_k}{{\bf T}_{ij}^2}
 \left[ \tilde{V}_{ij,k}\left(\tilde{p}_{ij},p_i,p_j,p_k,\phi_i,\phi_j\right) \right]_{\xi \xi'}
 \;\;
 {\cal A}_{n}^{\xi' \;(0)}\left(...,\tilde{p}_{ij},...,\tilde{p}_{k},...\right).
 \nonumber 
\eq
The functions $\tilde{V}_{ij,k}$ are given for the various splittings by
\bq
 \tilde{V}_{q_i g_j,k}\left(p,p_i,p_j,p_k,\phi_i,\phi_j\right)
 & = & 
 C_F
 {\cal R}
 \left[
       P_{q \rightarrow q g}\left(p,p_i,p_j,\phi_i,\phi_j\right)
     + 
       S_{q \rightarrow q g}\left(p,p_i,p_j,p_k,\phi_i,\phi_j\right)
 \right],
 \nonumber \\
 \tilde{V}_{g_i g_j,k}\left(p,p_i,p_j,p_k,\phi_i,\phi_j\right)
 & = & 
 C_A
 {\cal R}
 \left[
       P_{g \rightarrow g g}\left(p,p_i,p_j,\phi_i,\phi_j\right)
     + 
       S_{g \rightarrow g g}\left(p,p_i,p_j,p_k,\phi_i,\phi_j\right)
 \right],
 \nonumber \\
 \tilde{V}_{q_i \bar{q}_j,k}\left(p,p_i,p_j,p_k,\phi_i,\phi_j\right)
 & = & 
 T_R
 {\cal R}
 \left[
       P_{g \rightarrow q \bar{q}}\left(p,p_i,p_j,\phi_i,\phi_j\right)
 \right].
\eq
The operation ${\cal R}$ is defined in eq.~(\ref{def_R_operation}).
The mapped momenta $\tilde{p}_{ij}$ and $\tilde{p}_k$ are defined in the massless case by
\bq
\label{map_massless}
 \tilde{p}_{ij} = p_i + p_j - \frac{y}{1-y} p_k,
 \;\;\;\;\;\;
 \tilde{p}_k = \frac{1}{1-y} p_k,
 \;\;\;\;\;\;
 y = \frac{p_i \cdot p_j}{p_i \cdot p_j + p_i \cdot p_k + p_j \cdot p_k}.
\eq
In the massive case we use
\bq
\label{map_massive}
 \tilde{p}_k
 & = & 
 \frac{\sqrt{\lambda(Q^2,m_{ij}^2,m_k^2)}}{\sqrt{\lambda(Q^2,(p_i+p_j)^2,m_k^2)}}
 \left( p_k - \frac{Q \cdot p_k}{Q^2} Q \right) 
 + \frac{Q^2+m_k^2-m_{ij}^2}{2Q^2} Q,
 \nonumber \\
 \tilde{p}_{ij} & = & Q - p_k,
\eq
where $Q=p_i+p_j+p_k$ and $\lambda$ is the K\"allen function
\bq
 \lambda(x,y,z) & = & x^2 + y^2 + z^2 - 2 x y - 2 y z - 2 z x.
\eq
Eq.~(\ref{map_massive}) reduces in the massless limit to eq.~(\ref{map_massless}).

\subsubsection{Final-state emitter and initial-state spectator}

If the emitter is in the final state and the spectator in the initial state, 
the additional subtraction terms are given by
\bq
\lefteqn{
\tilde{\cal D}_{ij}^a
 = 
 - 4 \pi \alpha_s \mu^{2\eps} 
 } & & 
 \\
 & &
 {{\cal A}_{n}^{\xi \;(0)}}\left(...,\tilde{p}_{ij},...,\tilde{p}_{a},...\right)^\ast
 \;\;
 \frac{{\bf T}_{ij} \cdot {\bf T}_a}{{\bf T}_{ij}^2}
 \left[ \tilde{V}_{ij}^a\left(\tilde{p}_{ij},p_i,p_j,p_a,\phi_i,\phi_j\right) \right]_{\xi \xi'}
 \;\;
 {\cal A}_{n}^{\xi' \;(0)}\left(...,\tilde{p}_{ij},...,\tilde{p}_{a},...\right).
 \nonumber 
\eq
The dipole splitting function is related by crossing to the final-final case:
\bq
 \tilde{V}_{ij}^a\left(\tilde{p}_{ij},p_i,p_j,p_a,\phi_i,\phi_j\right)
 & = &
 \tilde{V}_{ij,a}\left(\tilde{p}_{ij},p_i,p_j,-p_a,\phi_i,\phi_j\right).
\eq
The mapped momenta $\tilde{p}_{ij}$ and $\tilde{p}_a$ are defined by
\bq
 \tilde{p}_{ij} = p_i + p_j - (1-x) p_a,
 \;\;\;\;\;\;
 \tilde{p}_a = x p_a.
\eq
The variable $x$ is given by
\bq
 x = \frac{p_i \cdot p_a + p_j \cdot p_a - p_i \cdot p_j + \frac{1}{2}\left(m_{ij}^2-m_i^2-m_j^2\right)}{p_i \cdot p_a + p_j \cdot p_a},
\eq
and reduces in the massless case to
\bq
 x & = & \frac{p_i \cdot p_a + p_j \cdot p_a - p_i \cdot p_j}{p_i \cdot p_a + p_j \cdot p_a}.
\eq

\subsubsection{Initial-state emitter and final-state spectator}

If the emitter is in the initial state and the spectator in the final state, 
the additional subtraction terms are given by
\bq
\lefteqn{
\tilde{\cal D}_{k}^{aj}
 = 
 - 4 \pi \alpha_s \mu^{2\eps} 
 } & & 
 \\
 & &
 {{\cal A}_{n}^{\xi \;(0)}}\left(...,\tilde{p}_{aj},...,\tilde{p}_{k},...\right)^\ast
 \;\;
 \frac{{\bf T}_{aj} \cdot {\bf T}_k}{{\bf T}_{aj}^2}
 \left[ \tilde{V}_{k}^{aj}\left(\tilde{p}_{aj},p_a,p_j,p_k,\phi_a,\phi_j\right) \right]_{\xi \xi'}
 \;\;
 {\cal A}_{n}^{\xi' \;(0)}\left(...,\tilde{p}_{aj},...,\tilde{p}_{k},...\right).
 \nonumber 
\eq
The dipole splitting function is related by crossing to the final-final case:
\bq
 \tilde{V}_{k}^{aj}\left(\tilde{p}_{aj},p_a,p_j,p_k,\phi_a,\phi_j\right)
 & = &
 {\cal C}_{(a,aj)} \tilde{V}_{aj,k}\left(-\tilde{p}_{aj},-p_a,p_j,p_k,\phi_a,\phi_j\right).
\eq
The operation ${\cal C}$ is defined in eq.~(\ref{def_C_operation}).
The mapped momenta $\tilde{p}_{aj}$ and $\tilde{p}_k$ are defined by
\bq
 \tilde{p}_{aj} = x p_a,
 \;\;\;\;\;\;
 \tilde{p}_k = p_k + p_i - (1-x) p_a,
 \;\;\;\;\;\;
 x = \frac{p_k \cdot p_a + p_i \cdot p_a - p_i \cdot p_k}{p_k \cdot p_a + p_i \cdot p_a}.
\eq
Note that we restrict ourselves to massless initial-state particles. This implies that the
masses of the particles $a$, $(aj)$ and $j$ are zero.

\subsubsection{Initial-state emitter and initial-state spectator}

If both the emitter and the spectator are in the initial state, 
the additional subtraction terms are given by
\bq
\lefteqn{
\tilde{\cal D}^{aj,b}
 = 
 - 4 \pi \alpha_s \mu^{2\eps} 
 } & & 
 \\
 & &
 {{\cal A}_{n}^{\xi \;(0)}}\left(\tilde{p}_1,...,\tilde{p}_{aj},...\right)^\ast
 \;\;
 \frac{{\bf T}_{aj} \cdot {\bf T}_b}{{\bf T}_{aj}^2}
 \left[ \tilde{V}^{aj,b}\left(\tilde{p}_{aj},p_a,p_j,p_b,\phi_a,\phi_j\right) \right]_{\xi \xi'}
 \;\;
 {\cal A}_{n}^{\xi' \;(0)}\left(\tilde{p}_1,...,\tilde{p}_{aj},...\right).
 \nonumber 
\eq
The dipole splitting function is related by crossing to the final-final case:
\bq
 \tilde{V}^{aj,b}\left(\tilde{p}_{aj},p_a,p_j,p_b,\phi_a,\phi_j\right)
 & = &
 {\cal C}_{a,aj} \tilde{V}_{aj,b}\left(-\tilde{p}_{aj},-p_a,p_j,-p_b,\phi_a,\phi_j\right).
\eq
In this case the mapped momenta are defined as follows:
\bq
 \tilde{p}_{aj} = x p_a,
 \;\;\;\;\;\;
 \tilde{p}_b = p_b,
 \;\;\;\;\;\;
 x = \frac{p_a \cdot p_b - p_i \cdot p_a - p_i \cdot p_b}{p_a \cdot p_b},
\eq
and all final state momenta are transformed as
\bq
 \tilde{p}_l & = & \Lambda p_l,
\eq
where $\Lambda$ is a Lorentz transformation defined by
\bq
 \Lambda^\mu_{\;\;\nu}
 & = & 
 g^\mu_{\;\;\nu}
 -2 \frac{\left(K^\mu+\tilde{K}^\mu\right) \left(K_\nu+\tilde{K}_\nu\right)}{\left(K+\tilde{K}\right)^2}
 + 2 \frac{\tilde{K}^\mu K_\nu}{K^2},
 \nonumber \\
 & &
 K = p_a + p_b - p_j,
 \;\;\;\;\;\;
 \tilde{K} = \tilde{p}_{aj} + p_b.
\eq
Again we consider only the case of massless initial-state particles. 
Therefore the 
masses of the particles $a$, $b$, $(aj)$ and $j$ are zero.

\section{Conclusions}
\label{sec:conclusions}

In this paper we extended the subtraction method for NLO calculations to random polarisations of the external particles.
We therefore may replace in the computation of the real contribution the sum over the helicity amplitudes by a smooth integration over
helicity angles.
We have derived the required additional subtraction terms, such that they match locally the singular behaviour of the squared amplitude with
random polarisations.
These additional subtraction terms integrate to zero and modify only the unintegrated subtraction term.
They can be used on top of any subtraction scheme, which provides local subtraction terms for the spin-summed amplitude squared.
Furthermore they are independent of any explicit definition of polarisation vectors or polarisation spinors.

\subsection*{Acknowledgements}

We would like to thank A. Kabelschacht for useful discussions on this subject.


\begin{appendix}

\section{Polarisation vectors and polarisation spinors}
\label{appendix:spinors}

We define the light-cone coordinates as
\bq
p_+ = p_0 + p_3, \;\;\; p_- = p_0 - p_3, \;\;\; p_{\bot} = p_1 + i p_2,
                                         \;\;\; p_{\bot^\ast} = p_1 - i p_2.
\eq
In terms of the light-cone components of a light-like four-vector, the corresponding massless spinors $\langle p \pm |$ and $| p \pm \rangle$ 
can be chosen as
\bq
\left| p+ \right\rangle = \frac{e^{-i \frac{\phi}{2}}}{\sqrt{\left| p_+ \right|}} \left( \begin{array}{c}
  -p_{\bot^\ast} \\ p_+ \end{array} \right),
 & &
\left| p- \right\rangle = \frac{e^{-i \frac{\phi}{2}}}{\sqrt{\left| p_+ \right|}} \left( \begin{array}{c}
  p_+ \\ p_\bot \end{array} \right),
 \nonumber \\
\left\langle p+ \right| = \frac{e^{-i \frac{\phi}{2}}}{\sqrt{\left| p_+ \right|}} 
 \left( -p_\bot, p_+ \right),
 & &
\left\langle p- \right| = \frac{e^{-i \frac{\phi}{2}}}{\sqrt{\left| p_+ \right|}} 
 \left( p_+, p_{\bot^\ast} \right),
\eq
where the phase $\phi$ is given by
\bq
p_+ & = & \left| p_+ \right| e^{i\phi},
 \;\;\;\;\;
 0 \le \phi < 2 \pi.
\eq
If the Cartesian coordinates $p_0$, $p_1$, $p_2$ and $p_3$ are real numbers, we have
\bq
 \left| p \pm \right\rangle^\dagger = e^{i \phi} \left\langle p \pm \right|,
 & &
 \left\langle p \pm \right|^\dagger = e^{i \phi} \left| p \pm \right\rangle,
 \;\;\;\;\;\;
 e^{i \phi} = \pm 1.
\eq
Spinor products are denoted as
\bq
 \langle p q \rangle = \langle p - | q + \rangle,
 & &
 [ q p ] = \langle q + | p - \rangle.
\eq 
Let $q$ be a light-like four-vector.
We define polarisation vectors for the gluons by
\bq
\eps_{\mu}^{+} = \frac{\langle q-|\sigma_{\mu}|p-\rangle}{\sqrt{2} \langle q p \rangle},
 & &
\eps_{\mu}^{-} = \frac{\langle q+|\bar{\sigma}_{\mu}|p+\rangle}{\sqrt{2} [ p q ]},
\eq
with $\sigma_\mu = ( 1, \vec{\sigma} )$ and $\bar{\sigma}_\mu = ( 1, - \vec{\sigma} )$, where $\vec{\sigma}=(\sigma_1,\sigma_2,\sigma_3)$ are the
Pauli matrices.
The dependence on the reference four-vector $q$ drops out in gauge invariant quantities.
In numerical calculations a convenient choice \cite{Kleiss:1988ne}
for the reference four-vector $q$ for a gluon with momentum $p_\mu=(p_0,p_1,p_2,p_3)$
is given by $q_\mu = (p_0,-p_1,-p_2,-p_3)$.
With this choice the spinor products $\langle q p \rangle$ and $[p q]$ in the denominator of the 
polarisation vectors vanish only in the soft limit. In the soft limit the numerator vanishes at the same rate.
This ensures that the evaluation of the polarisation vector is always stable.
Under complex conjugation we have
\bq
 \left( \eps_\mu^+ \right)^\ast = \eps_\mu^-,
 & &
 \left( \eps_\mu^- \right)^\ast = \eps_\mu^+.
\eq
The reference four-vector $q$ can be used to project any not necessarily light-like four-vector $P$ 
on a light-like four-vector $P^\flat$:
\bq
\label{projection_null}
 P^\flat & = & P - \frac{P^2}{2 P \cdot q} q.
\eq
The four-vector $P^\flat$ satisfies $(P^\flat)^2=0$.
Let $P$ be a four-vector satisfying $P^2=m^2$. We define the spinors associated to massive fermions by
\bq
\label{eq:spinors} 
 u^\pm = \frac{1}{\langle P^\flat \pm | q \mp \rangle} \left( P\!\!\!/ + m \right) | q \mp \rangle,
 & &
\bar{u}^\pm = \frac{1}{\langle q \mp | P^\flat \pm \rangle} \langle q \mp | \left( P\!\!\!/ + m \right),
 \nonumber \\
 v^\mp = \frac{1}{\langle P^\flat \pm | q \mp \rangle} \left( P\!\!\!/ - m \right) | q \mp \rangle,
 & &
\bar{v}^\mp = \frac{1}{\langle q \mp | P^\flat \pm \rangle} \langle q \mp | \left( P\!\!\!/ - m \right).
\eq
These spinors satisfy the Dirac equations
\bq 
\left( p\!\!\!/ - m \right) u^\lambda = 0, & & 
\bar{u}^\lambda \left( p\!\!\!/ - m \right) = 0,
 \nonumber \\
\left( p\!\!\!/ + m \right) v^\lambda = 0, & & 
\bar{v}^\lambda \left( p\!\!\!/ + m \right) = 0,
\eq
the orthogonality relations
\bq
\bar{u}^{\bar{\lambda}} u^\lambda = 2 m \delta^{\bar{\lambda}\lambda}, 
 & &
\bar{v}^{\bar{\lambda}} v^\lambda = -2 m \delta^{\bar{\lambda}\lambda}, 
\eq
and the completeness relation
\bq
\sum\limits_{\lambda} u^\lambda \bar{u}^\lambda = p\!\!\!/ + m,
 & &
\sum\limits_{\lambda} v^\lambda \bar{v}^\lambda = p\!\!\!/ - m.
\eq
We further have
\bq
 \bar{u}^{\bar{\lambda}} \gamma^\mu u^\lambda & = & 2 p^\mu \delta^{\bar{\lambda} \lambda},
 \nonumber \\
 \bar{v}^{\bar{\lambda}} \gamma^\mu v^\lambda & = & 2 p^\mu \delta^{\bar{\lambda} \lambda}.
\eq
In the massless limit the definition reduces to
\bq
 \bar{u}^\pm = \bar{v}^\mp = \langle p \pm |, 
 & &
 u^\pm = v^\mp = | p \pm \rangle.
\eq
Let us denote the helicity projection operators by
\bq
 P_+ & = & \frac{1}{2} \left( 1 + \gamma_5 \right) = 
 \left( \begin{array}{cc} 1 & 0 \\ 0 & 0 \\ \end{array} \right),
 \nonumber \\
 P_- & = & \frac{1}{2} \left( 1 - \gamma_5 \right) =  
 \left( \begin{array}{cc} 0 & 0 \\ 0 & 1 \\ \end{array} \right).
\eq
Then
\bq
 \left| p+ \right\rangle \left\langle p+ \right| = P_+ p\!\!\!/,
 & &
 \left| p- \right\rangle \left\langle p- \right| = P_- p\!\!\!/,
\eq
The combinations $\left| p+ \right\rangle \left\langle p- \right|$ and
$\left| p- \right\rangle \left\langle p+ \right|$ are a little bit more complicated:
\bq
 \left| p+ \right\rangle \left\langle p- \right| 
 & = &
 \frac{1}{\sqrt{2}} P_+ p\!\!\!/ \eps\!\!\!/^-(p,q)
 =
 - \frac{1}{\sqrt{2}} P_+ \eps\!\!\!/^-(p,q) p\!\!\!/,
 \nonumber \\
 \left| p- \right\rangle \left\langle p+ \right| 
 & = &
 - \frac{1}{\sqrt{2}} P_- p\!\!\!/ \eps\!\!\!/^+(p,q)
 =
 \frac{1}{\sqrt{2}} P_- \eps\!\!\!/^+(p,q) p\!\!\!/.
\eq

\end{appendix}

\bibliography{/home/stefanw/notes/biblio}
\bibliographystyle{/home/stefanw/latex-style/h-physrev5}

\end{document}